\documentclass[twocolumn,amsmath,amssymb,prb,superscriptaddress,reprint]{revtex4}
\pdfoutput=1
\usepackage{amsmath}
\usepackage{amssymb}
\usepackage{longtable}
\usepackage[dvips]{graphicx}
\usepackage{setspace}
\usepackage{color} 
\usepackage{moreverb}
\usepackage{subfigure}
\usepackage{bbm}
\begin{document}
\title{Theoretical Comparison of Rashba Spin-Orbit Coupling in Digitally, Discretely, and Continuously Alloyed Nanostructures}
\author{Joseph Pingenot}
\affiliation{Center for Semiconductor Physics in Nanostructures, The
  University of Oklahoma, Norman, OK 73019}
\author{Kieran Mullen}
\affiliation{Center for Semiconductor Physics in Nanostructures, The
  University of Oklahoma, Norman, OK 73019}
\date{\today}

\begin{abstract} Although most theoretical calculations of quantum wells with non-square profiles assume that material composition is varied continuously, it is more common in experiment to grow digital alloys.  We compare the Rashba spin-orbit
interaction of triangular wells using continuous, discrete, and
digital alloying profiles in (001)-grown
triangular InSb/Al$_{f(z)}$In$_{1-f(z)}\text{Sb}$, finding a very large difference between digital alloying and the others, including a sign change in the Rashba spin-orbit coupling.  We find that the interface
contribution to the Rashba spin-orbit coupling is much larger in the
continuously- and discretely-alloyed triangular quantum wells than in
the digitally-alloyed triangular wells, in which it is almost completely absent.  The electric field
contribution, however, is quite similar in all three systems.  Due to
a much stronger doping dependence in all three systems, the electric
field contribution dominates at higher dopings, although the very large offset
due to the near absence of interface contribution in
digitally-alloyed wells persists.
\end{abstract} 
\maketitle

The spin-orbit interaction couples the spin of an electron with the
real-space wavefunction, adding a $\vec k \times \vec
\sigma$-dependent term to the Hamiltonian.  This serves as the basis
for a good deal of very interesting physics, such as topological
insulators\cite{Hasan.RMP.2010} and weak
antilocalization\cite{Koga.PRL.2002,Zhou.APL.2008,Nitta.JAP.2009}.  In
semiconductor nanostructures, band structure brings about two specific
spin-orbit couplings, the Rashba and Dresselhaus interactions.  Unlike
the Dresselhaus spin-orbit interaction, which arises from inversion
asymmetry in the bulk materials, the Rashba interaction arises from
inversion asymmetry of the structure.  This allows the Rashba
spin-orbit interaction to be controlled via applied electric fields,
and thus presents a means for coherent control of the electron spin by
electrical gates.  Coherent control of the electron spin through the
Rasbha spin-orbit coupling is an important consideration in quantum
computing with spins in quantum dots\cite{Stepanenko.PRL.2004}, and
may also provide the basis for novel devices, such as the Datta-Das
spin transistor\cite{Datta1990}.

\begin{figure}[h!c]
\begin{center}
\subfigure{\includegraphics[width=0.49
    \columnwidth]{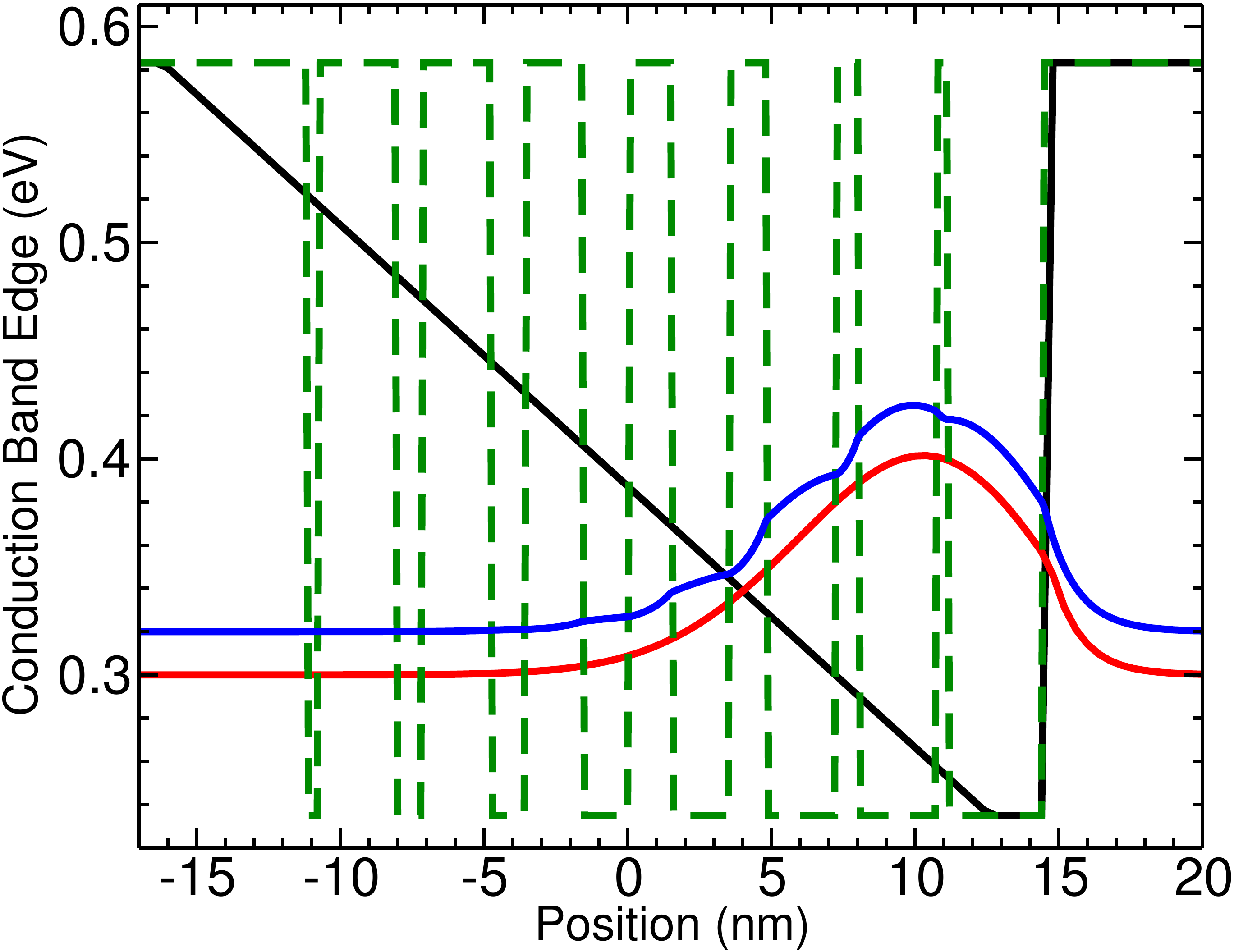}\label{sfig:InSb_wf_alpha}}
\hfill
\subfigure{\includegraphics[width=0.49
    \columnwidth]{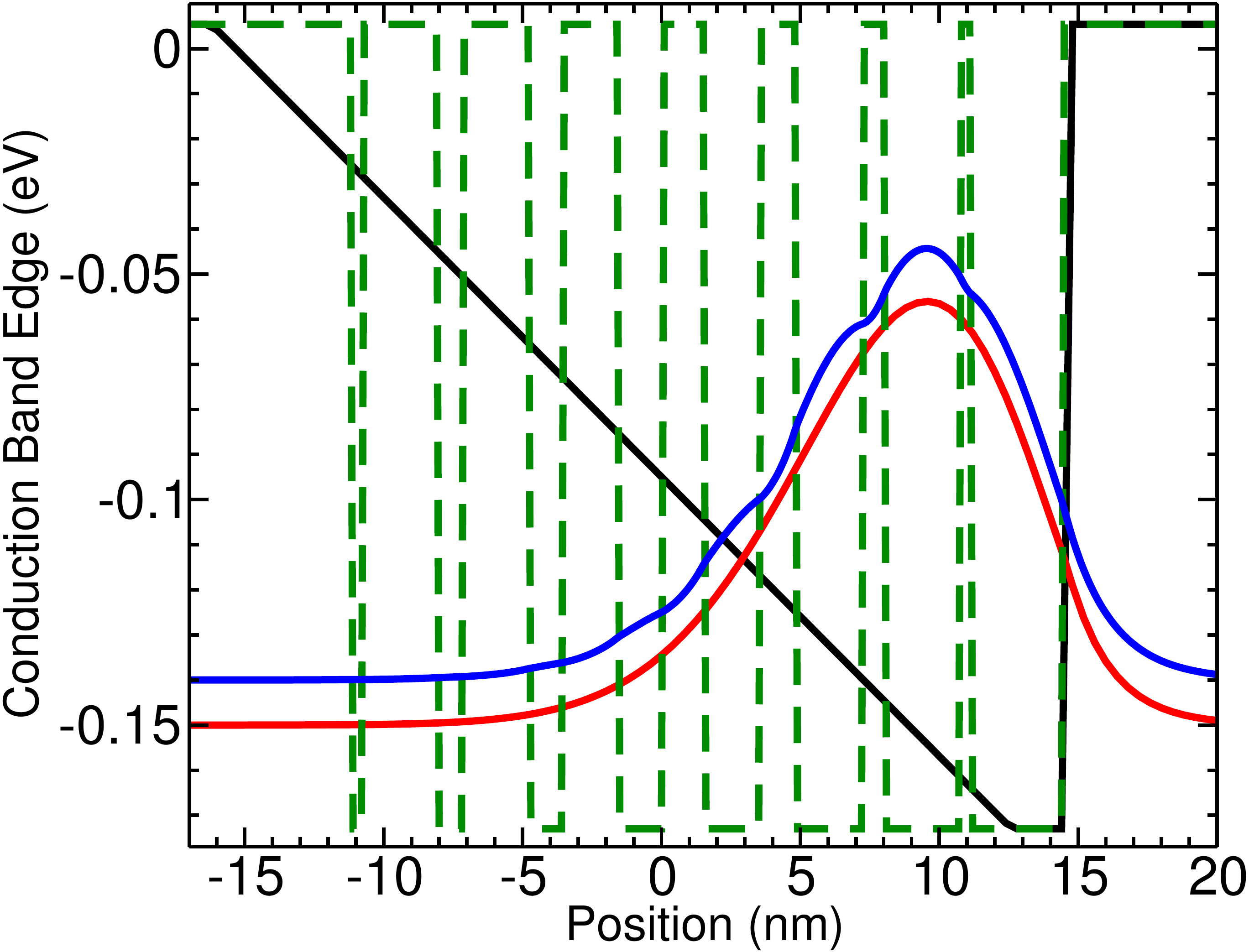}\label{sfig:gainas_wf_profile}}
\caption{ Continuously- and digitally-alloyed well conduction band
  profiles (solid and dashed lines, respectively) and the resulting
  ground states (bottom and top solid lines).  Left plot is
  InSb/$\text{Al}_{0.2}\text{In}_{0.8}\text{Sb}$; right is
  InAs/$\text{Ga}_{0.2}\text{In}_{0.8}\text{As}$.  The
  first-derivative discontinuities in the InSb wavefunction are due to
  an inhomogeneous effective mass, but do not have a large effect on
  the result; in the InAs well, the effective masses are much closer.
  (color online)\label{fig:wavefunctions} }
\end{center}
\end{figure}

To grow $(001)$ quantum wells of the form
$\text{A}_{f(z)}\text{B}_{1-f(z)}$, where $f(z)$ need not be constant,
three primary methods can be used.  The first,
continuous alloying, recreates the function $f(z)$ by
continuously controlling shutter aperture and cell temperature in the
molecular beam epitaxy (MBE) system.  It is very difficult to 
tune the parameters during growth to reproduce $f(z)$ accurately,
however.  An example of a continuously-alloyed triangular well conduction band profile is shown with black solid lines in figure \ref{fig:wavefunctions}.

It is easier to approximate $f(z)$ by a sequence of steps which
average $f(z)$ across each step.  Growth is then stopped between each
step, the cell temperatures are allowed to stabilize, and then growth
continues for the step.  This method, discrete alloying, requires much
care, and so digital alloying is often used.

A digitally-alloyed nanostructure takes discrete alloying to the next level and
builds a nanostructure by creating a sequence of square quantum
wells.\cite{Agarwal.ECS.2010,Offermans.2003} The steps are then broken
into well and barrier regions, with the percentage of well or barrier
corresponding to the value of $f(z)$ averaged across the step.  An example of a digitally-alloyed triangular well conduction band profile is shown using green dashed lines in figure \ref{fig:wavefunctions}.

Figure \ref{fig:wavefunctions} compares continuously and digitally
alloyed wells and their ground state wavefunctions (from a self-consistent
calculation described below) in InSb/InAlSb and InAs/InGaAs triangular
quantum wells.  The wavefunctions are very similar, save for some
oscillations in the digital alloy.  Because of the close similarity in
wavefunctions and greater ease of well growth, digital alloying is
often used to grow non-constant $f(z)$ wells instead of the discrete
or continuous alloying methods.

In this paper, we compare the Rashba spin-orbit interaction of triangular wells using continuous, discrete, and digital alloying profiles.  We find that, although the electric field component of the Rashba spin-orbit coupling and ground state wavefunction are similar between continuously-, discretely-, and digitally-alloyed quantum wells, the square well basis of the digitally-alloyed well causes the interface component of the Rashba coupling to be considerably smaller than is the case in the continuous and discrete wells.

The Rashba spin-orbit interaction has been shown \cite{Schapers1998,Lange.1996} to consist of two parts: an electric field contribution and a valence band offset contribution.  The electric field component can be understood as an average of the electric field by the effective mass wavefunction $\psi(z)$, weighted by a bandstructure-determined factor.  This contribution takes the form:
\begin{equation}\label{eqn:a_e}
\vec \alpha_E = \vec \alpha_0 \int\limits_{-\infty}^{+\infty}{\left|\psi(z)\right|^2\frac{d\phi(z)}{dz}
 \sum\limits_{b=7,8}{\frac{\left(-1\right)^{b+1}}{\left(E-E_{\Gamma_b}(z)-\phi(z)\right)^{2}}}}dz
\end{equation}
where $\vec \alpha_0 = \hat z \frac{\hbar^2 E_p}{6 m_0}$, $E_{\Gamma_b}(z)$ is the band edge of the $\Gamma_b$ band through the nanostructure, $\phi$ is the electrostatic potential energy from applied and internal self-consistent electric fields, $\psi(z)$ is the effective mass wavefunction, $E_p$ is the $k \cdot p$ interband coupling parameter, and $m_0$ is the bare mass of the electron.

The other contribution to the Rashba spin-orbit interaction comes from the valence band offsets at the quantum well interfaces.  At a perfectly sharp interface between two materials, these offsets form a Dirac delta function at the interface.  When integrated, this becomes the change in the valence bands across the interface and an average of the valence band weighting term across the interface.  This contribution takes the form:
\begin{equation}\label{eqn:a_i}
\vec \alpha_I = \vec \alpha_0 \sum\limits_{i=0}^{N_I} \sum\limits_{b=^7_8;s=_+^-} \frac{\left(-1\right)^{b+1}\left(E_{\Gamma_b}^+(z_i) - E_{\Gamma_b}^-(z_i)\right)}{2\left(E-E_{\Gamma_b}^s(z_i)-\phi(z_i)\right)^{2}}\left|\psi(z_i)\right|^2
\end{equation}
where $N_I$ is the number of interfaces, and $z_i$ is the location of the interface $i$.  The interface contribution can be understood as the difference between the {\it valence} band edges at each interface, weighted by the average inverse-square of the electrostatic-potential-offset band edge and the probability density at the interface.  In the case of a continuously-alloyed well, the single interface becomes a continuous sequence of infinitesimal changes.  When used on a real-space grid, as in this work, the material is constant within each gridsite.  This yields a discrete alloy with steps of one gridsite.

\begin{figure}[hh!c]
\begin{center}
  \includegraphics[width=0.5 \columnwidth]{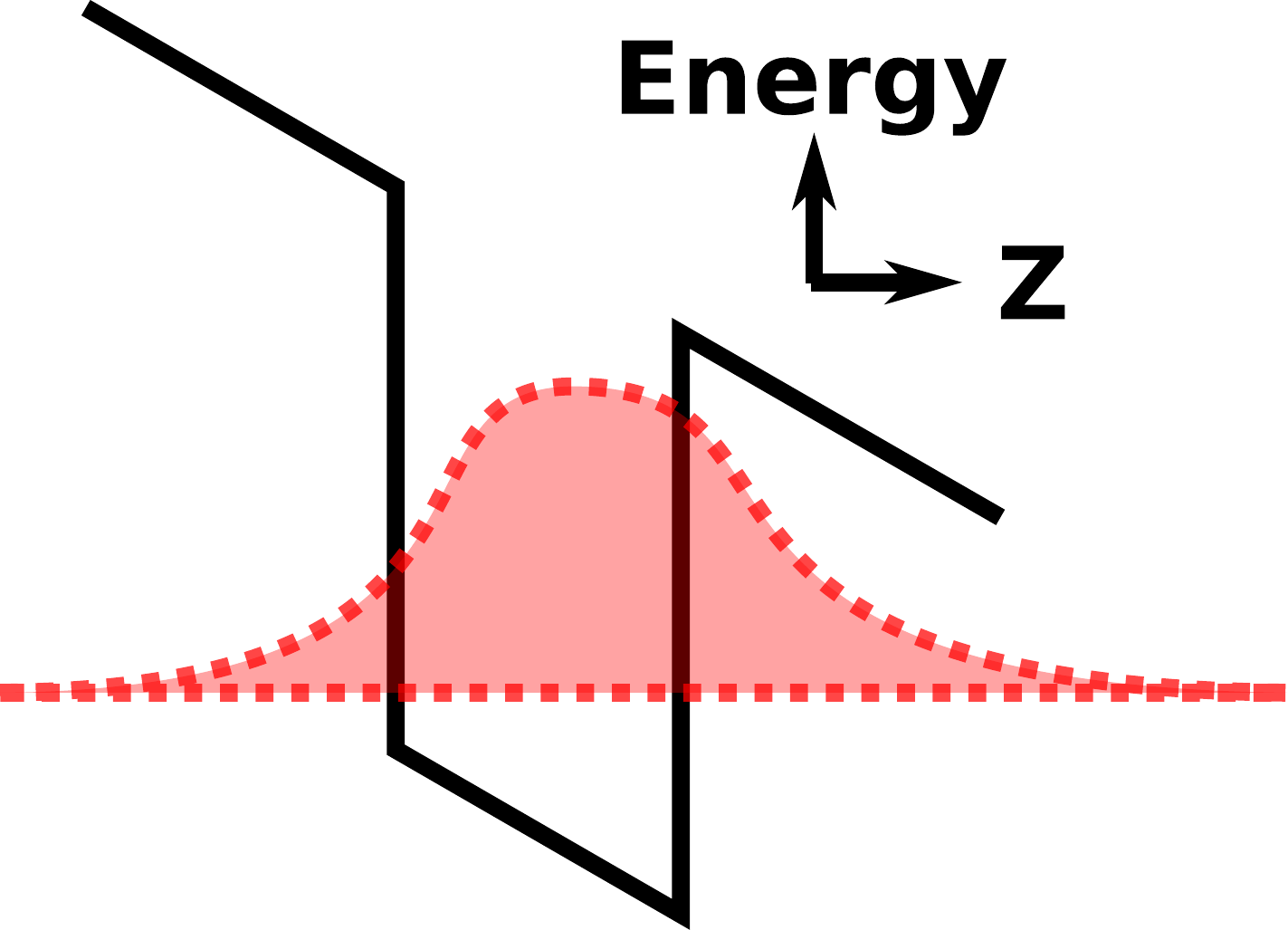}
    \caption{
      Cartoon of a wavefunction in the conduction band of a square well.  A constant electrostatic field $\phi(z) = kz$ has been applied to the structure.  Valence bands are not shown due to space considerations, but are also square.
    }
    \label{fig:square_well}
  \end{center}
\end{figure}

Examining a square well with a constant electric field (fig. \ref{fig:square_well}) can provide intuition.
In this square well, the valence band difference at the left and right interfaces are of equal magnitude but opposite sign.  The inverse-square band offset weighting terms are also the same on the left and the right, if we neglect the electric field term.  If a non-uniform electrostatic potential were introduced, for instance from the internal electric field between electrons in the well and the dopant layer,  a small difference would arise.  A much larger contribution comes from the difference in probability density at the left and right interfaces.  This will in general be the largest contribution to the interface term.

It should be noted that these equations do not account for Dirac renormalization\cite{Winkler2003}, and thus neglect e.g. the Darwin contribution to the overall Hamiltonian.  This is not a major impediment to the current discussion, as the Dirac renormalization terms have only a modest effect on the above equations.

Much more interesting is the energy $E$ above.  The effective mass wavefunction perturbation methods used to determine equations \ref{eqn:a_e} and \ref{eqn:a_i} are derived using a projection operator formalism which has been described by L\"owdin.\cite{Lowdin.JMP.1962}  Ultimately, the $E$ term in equations \ref{eqn:a_e} and \ref{eqn:a_i} comes from the solution to the full N-band Schr\"odinger equation for {\em all} of the carriers in the well.  Both aspects of this are important to fully understand the Rasbha spin-orbit interaction in doped quantum wells: the fact that these come from a solution to the full Schr\"odinger equation as well as the fact that it must include every carrier in the well.

One approach is to use the energy of the particular subband being
examined.  In this approach, the single-particle, effective mass
ground state eigenenergy ($E_0$) is used for $E$.  The drawback of
this approach is that it neglects the carriers at higher $k$, at
energies up to the Fermi energy.  This approach is taken by some
groups\cite{Esmerindo.PRL.2007}, although they approximate $E_0$ by
the energy of the conduction band edge in the well.  Although $E_0$ is
readily understood in triangular or more general quantum wells, the
value to take for the conduction band edge in triangular or more
complicated wells is unclear.

Alternately, a clear second approach is to take the Fermi energy ($E_F$) for $E$.  This then can be qualitatively understood as examining only the highest occupied state.  This approach has also been taken by a number of groups\cite{Engels.PRB.1997,Koga.PRL.2002,Nitta.JAP.2009}, and like the first approach, some\cite{Engels.PRB.1997} approximate $E_F$ with the conduction band edge in the well.  However, other groups use $E_F$.  In addition, using the Fermi energy instead of the conduction band edge in the well\cite{Zhou.APL.2008} should qualitatively reproduce the trend of decreasing Rashba spin-orbit coupling with increasing carrier density (i.e. increasing $E_F$) due to the inverse-$E^2$ dependency of $\alpha$, which is harder to explain with other means.

In this paper, we use the latter definition of $E$, namely that we
simply replace $E$ with $E_F$.  However, simply introducing $E_F$ is
somewhat ad hoc and a more rigorous formalism is certainly called for
to clarify the ambiguity in methods.  Koga et al.\cite{Koga.PRL.2002}
compared the approximation for $E$ used by Engels et
al.\cite{Engels.PRB.1997} (the conduction band edge in the well) and
found approximately a 20-30\% increase in $\alpha$ when using $E_F$
instead of the band edge in InAlAs/InGaAs/InAlAs wells (70-90 meV).
The ground state energy, necessarily being lower than $E_F$ and higher
than $E_0$ should lie somewhere below that.  The difference in
InSb/InAlSb is likely even larger, due to the smaller bandgap in InSb.
Using the $E_0$ or conduction band edge approximation of $E$ should
change only the scale of our results, not the results themselves.

To compare digitally-, discretely-, and continuously-alloyed triangular wells, we have calculated the Rashba spin-orbit coupling in doped continuously-alloyed, discretely-alloyed, and digitally-alloyed quantum wells without an applied electric field.  The wavefunctions within the conduction band were calculated using a real-space self-consistent effective mass method, using an inhomogeneous, material-dependent effective mass.  As the abrupt change in effective mass between the well and barrier in InSb/AlInSb quantum wells causes discontinuities in the first derivative of the wavefunction, we additionally calculated the same well types but with an InAs/GaInAs quantum well system where the effective masses in the well and barrier are much closer.

To prevent infinite loops in the self-consistent calculation, both the
second-to-last and previous wavefunctions were compared to the current
wavefunction to determine convergence, with no potential mixing being
performed.  An alternate method would have been to mix a fraction of
the previous and current potential values.\cite{Ando.RMP.1982}

After the effective mass wavefunctions were calculated, equations \ref{eqn:a_e} and \ref{eqn:a_i} were used to calculate the Rashba spin-orbit coupling for the conduction electrons in the ground state subband.

\begin{figure}[h!c]
\begin{center}
\subfigure{\includegraphics[width=0.49 \columnwidth]{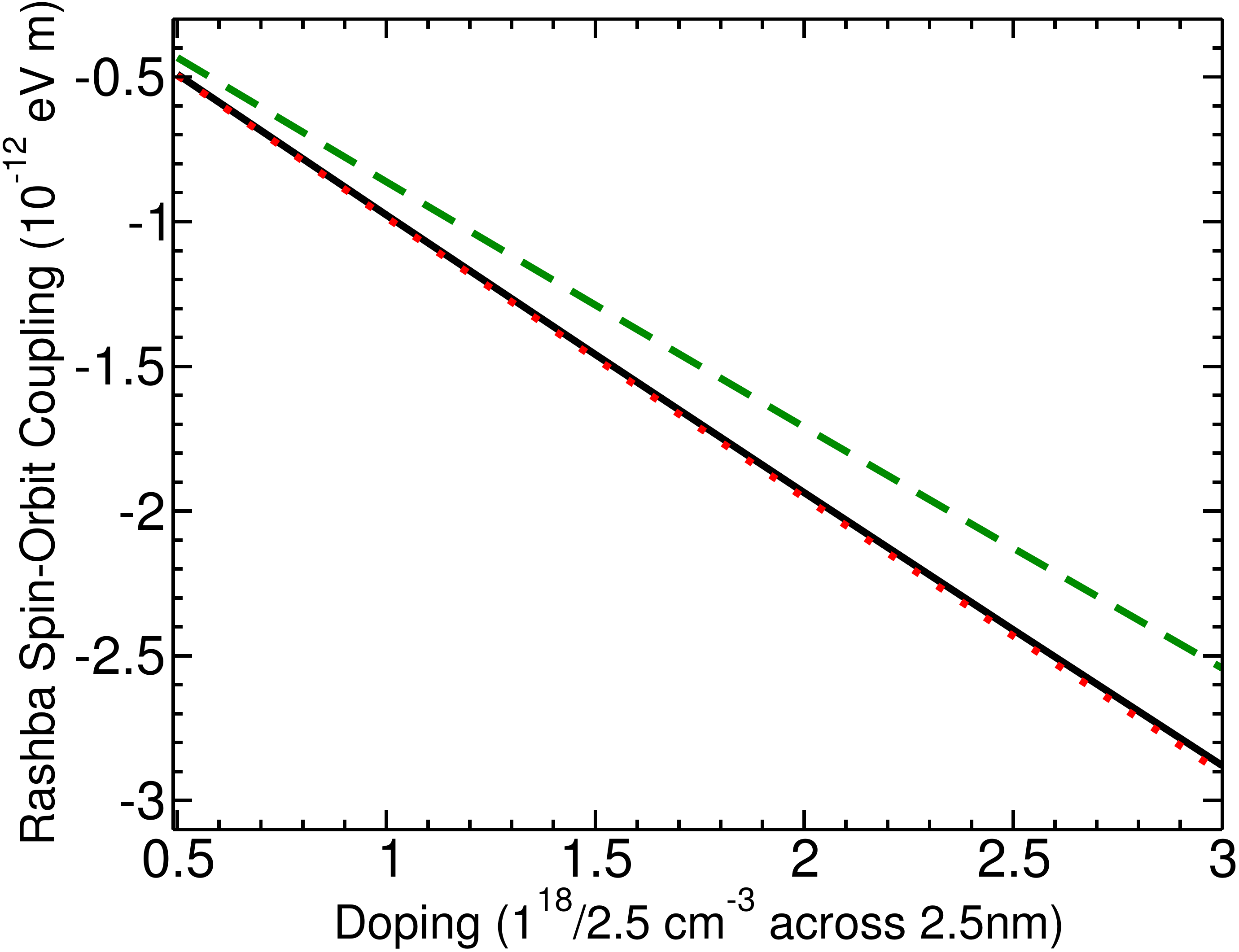}\label{sfig:insb_efield_alpha}}
\hfill
\subfigure{\includegraphics[width=0.49 \columnwidth]{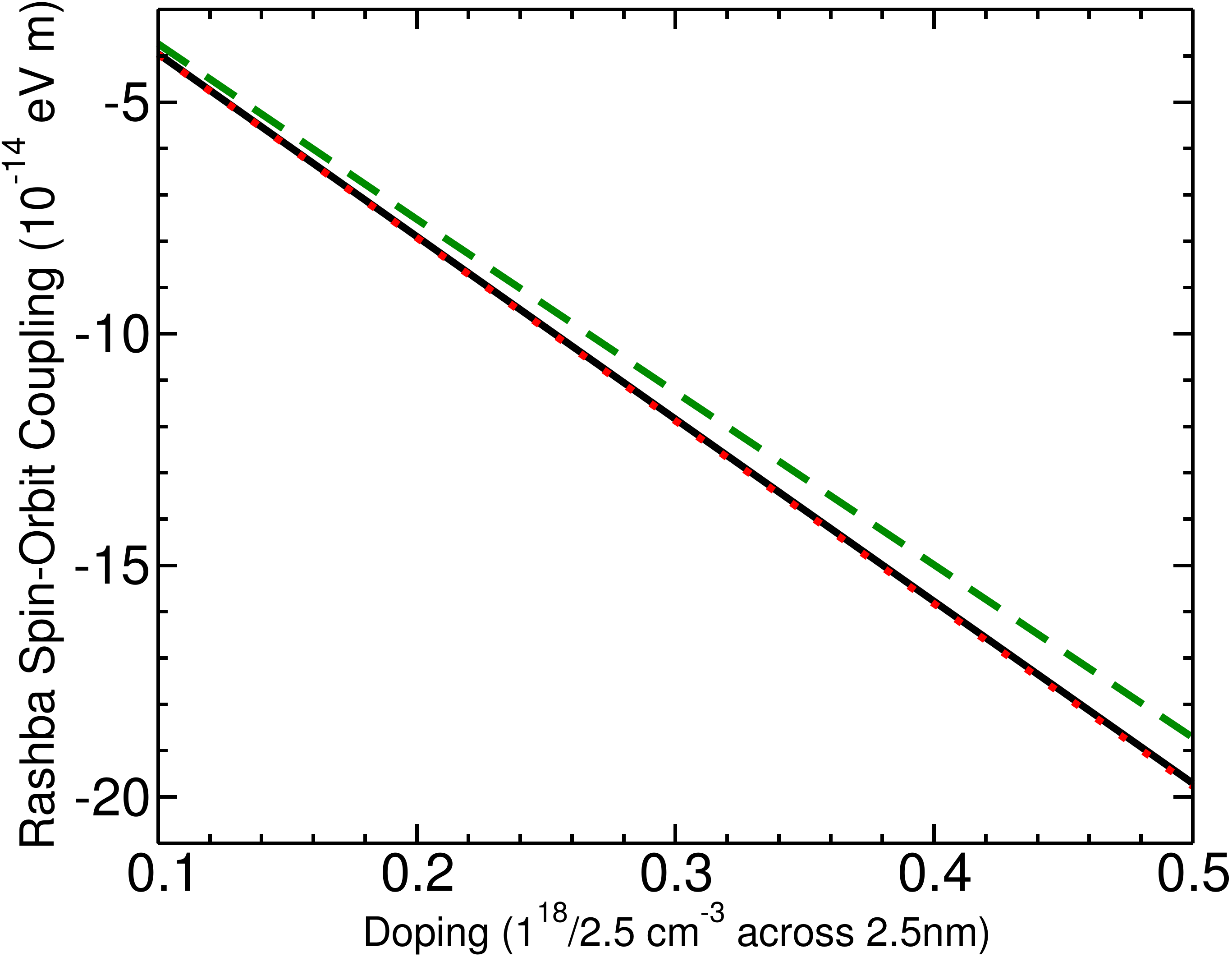}\label{sfig:gainas_efield_alpha}}
\caption{
  Electric field contribution to the Rashba spin-orbit coupling in InSb/$\text{Al}_{0.2}\text{In}_{0.8}\text{Sb}$ (left plot) and InAs/$\text{In}_{0.2}\text{Ga}_{0.8}\text{As}$ (right plot) continuously-alloyed (solid line), discretely-alloyed (dotted line), and digitally-alloyed (dashed line) triangular quantum wells. Results for discrete and continuous alloys are nearly indistinguishable.\label{fig:efield_alpha}
}
\end{center}
\end{figure}

\begin{figure}[h!c]
\begin{center}
\subfigure{\includegraphics[width=0.49 \columnwidth]{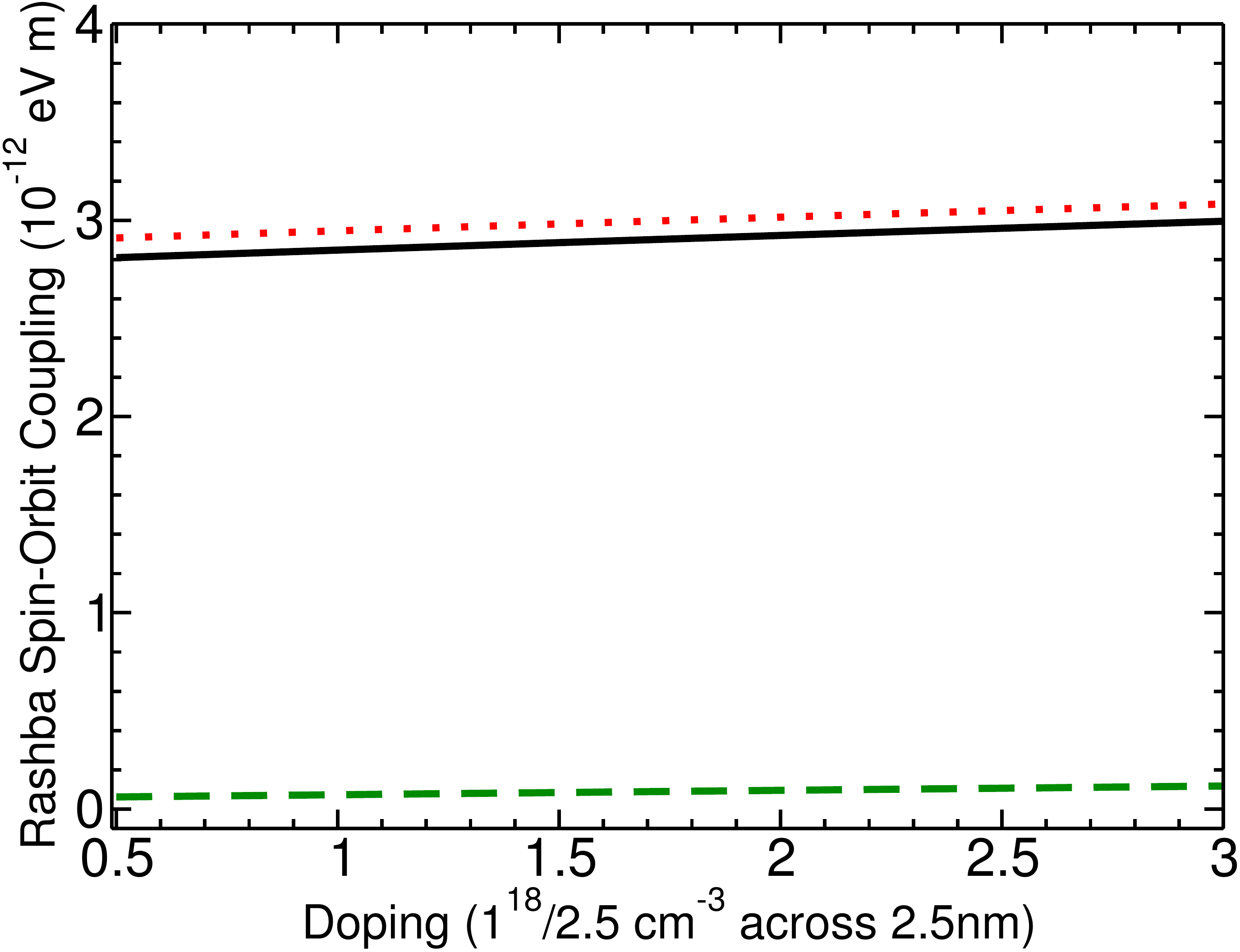}\label{sfig:insb_iface_alpha}}
\hfill
\subfigure{\includegraphics[width=0.49 \columnwidth]{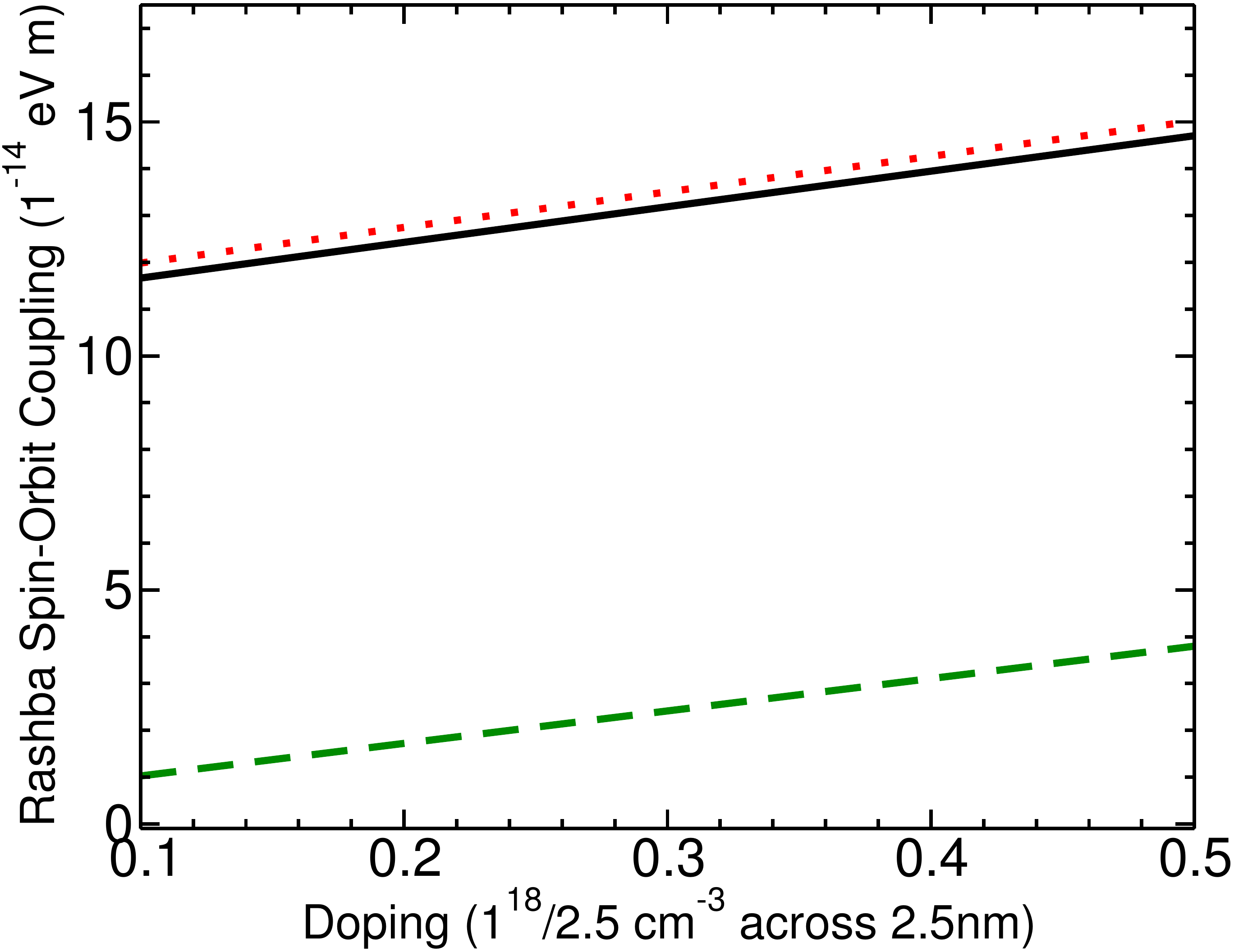}\label{sfig:gainas_iface_alpha}}
\caption{
  Interface contribution to the Rashba spin-orbit coupling in InSb/$\text{Al}_{0.2}\text{In}_{0.8}\text{Sb}$ (left plot) and InAs/$\text{In}_{0.2}\text{Ga}_{0.8}\text{As}$ (right plot) continuously-alloyed (solid line), discretely-alloyed (dotted line), and digitally-alloyed (dashed line) triangular quantum wells.\label{fig:interface_alpha}
}
\end{center}
\end{figure}

\begin{figure}[h!c]
\begin{center}
\subfigure{\includegraphics[width=0.49 \columnwidth]{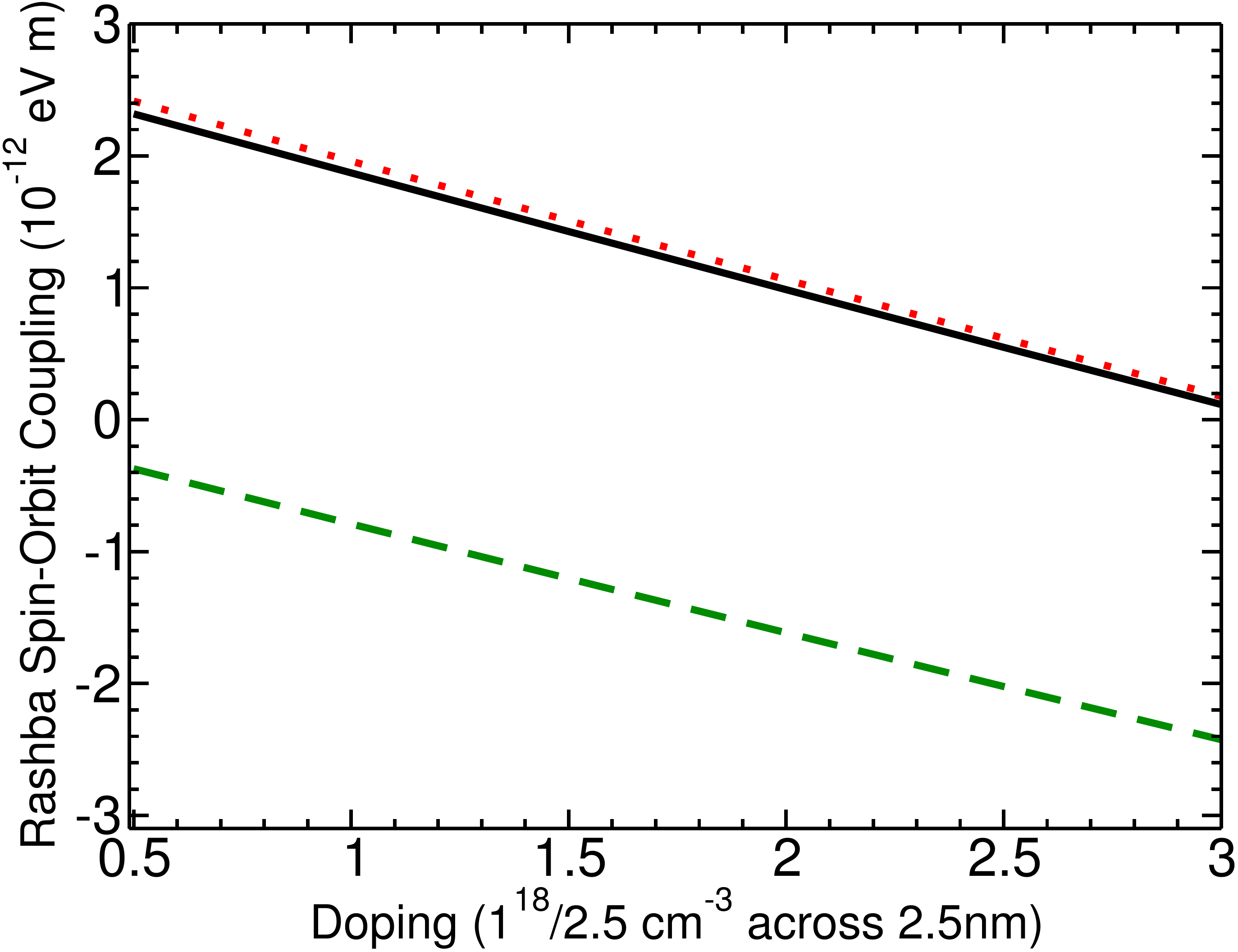}\label{sfig:insb_total_alpha}}
\hfill
\subfigure{\includegraphics[width=0.49 \columnwidth]{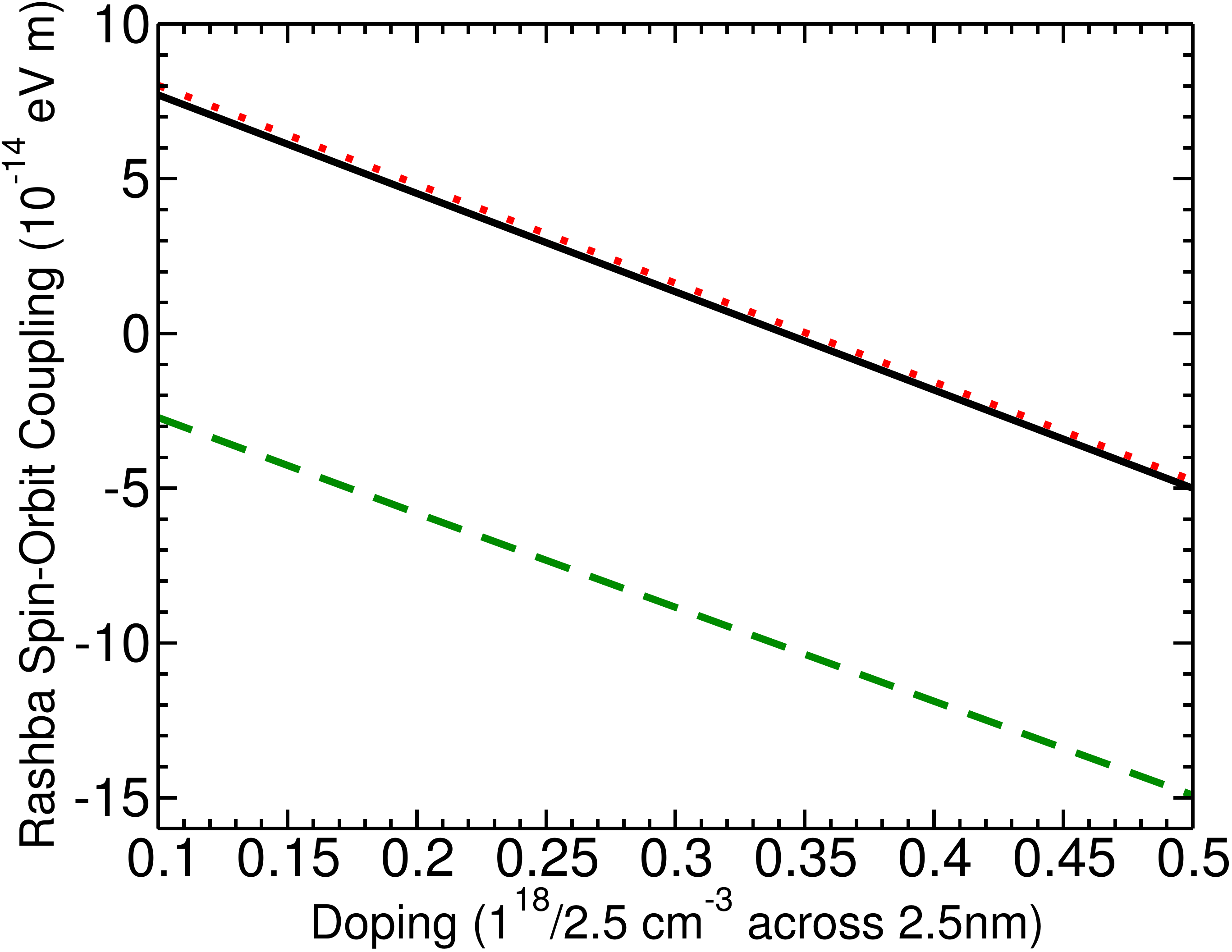}\label{sfig:gainas_total_alpha}}
\caption{
  Total Rashba spin-orbit coupling in InSb/$\text{Al}_{0.2}\text{In}_{0.8}\text{Sb}$ (left plot) and InAs/$\text{In}_{0.2}\text{Ga}_{0.8}\text{As}$ (right plot) continuously-alloyed (solid line), discretely-alloyed (dotted line), and digitally-alloyed (dashed line) triangular quantum wells. Continuous and discrete alloys are almost indistinguishable.\label{fig:total_alpha}
}
\end{center}
\end{figure}

The Rashba couplings at a variety of dopings were thus calculated and
are plotted in figure \ref{fig:total_alpha}.  The electric field and interface contributions are plotted in figures \ref{fig:efield_alpha} and \ref{fig:interface_alpha}, respectively.
Doping was varied between $2\times 10^{17}
\text{cm}^{-3}$ and $1.2\times 10^{18} \text{cm}^{-3}$
across 2.5nm in the InSb/$\text{In}_{f(z)}\text{Al}_{1-f(z)}$Sb
quantum wells, and between $4\times 10^{16}
\text{cm}^{-3}$ and $2\times 10^{17} \text{cm}^{-3}$
across 2.5nm in the InAs/$\text{In}_{f(z)}\text{Ga}_{1-f(z)}$As
quantum wells.  The doping regions in both cases began 15nm to the
right of the sharp interface of the triangular wells.  The wells were all 28.8nm wide, using a grid spacing of 0.4nm for the continuously and discretely alloyed wells and 0.1nm for the digitally alloyed well.

Due to the similarity in wavefunction shape, the electric field term can be expected to be very similar for digitally- and continuously-alloyed wells.  This expectation is borne out in our calculations, shown in subfigures \ref{sfig:insb_efield_alpha} and \ref{sfig:gainas_efield_alpha}.
There is a strong increase in the magnitude of the Rashba spin-orbit interaction as doping is increased for all three alloying types.  This is due to the increasing electric field between the carriers in the well and the dopant layer as doping is increased, and is expected.

In a continuously alloyed well, given that one side is a sharp interface, the other side of the well provides a continuous (or monotonically changing) set of interfaces.  In a digitally alloyed well, by contrast, the sequence of square wells present both a strong positive and negative interface at the subwell boundaries.  This can be seen in figure \ref{fig:wavefunctions} where the continuously-alloyed and digitally-alloyed wells and wavefunctions are superimposed (without the internal potential) for InSb and InAs triangular quantum wells.  Thus instead of one large change in the valence band fighting a large set of small changes, a digitally alloyed well has two large changes in opposition.  In addition, figure \ref{fig:wavefunctions} shows an increasing probability density in the first subwell (going right to left) which is then countered by a series of subwells with decreasing probability density.  Because the valence band changes are identical in each of the subwells, this increase in probability density in the first subwell followed by a decrease across the subsequent subwells seen in figure \ref{fig:wavefunctions} serves to counter itself in addition to the small changes within each individual subwell.

There is a moderate increase in the magnitude of the interface contribution to the Rashba spin-orbit interaction as doping is increased.  This has its origin in two parts: the first is the increase in electrostatic potential on one side of the well due to the internal electric field between the carriers in the well and the dopant layer, leading to a larger asymmetry across the wells in the valence band.  The second part is the wavefunction tunneling into the barrier increasing as the internal electric field increases, leading to another increase in the asymmetry of the well and therefore an increasing interfacial Rashba contribution.

The sum of the two contributions is plotted for the quantum wells in
figure \ref{fig:total_alpha}.  Because of the almost complete absence
of the interface contribution in the digital alloys but the similarity
of the electric field contribution to that found in discrete or
continuous quantum wells, the total Rashba spin-orbit interaction in
the digitally alloyed wells is considerably different from the
continuous and discrete counterparts.  This is perhaps surprising, due
to the similarity of digital and continuous alloy wavefunctions.  This
further reinforces the importance of the interface term in Rashba
calculations.

Aside from the offset due to the differences in interface terms, the
similarity in electric field terms in digital compared to discrete and
continuous alloying causes the high-doping regimes, where the electric
field term dominates, to have similar trends.  This has its origin in
the greater sensitivity of the electric field contribution to the
internal electric field compared to the interface contribution as
doping is increased.

To summarize, digitally alloyed quantum wells have substantially different Rashba spin-orbit couplings compared to continuously or discretely alloyed wells, even changing sign in some doping regimes.  This has its origin in the near absence of the interface contribution in digital alloys.  As the electric field contribution is more sensitive to doping than the interface contribution, the trend as a function of doping density in the high-doping regime in digital alloys are similar to the trend in continuously or discretely alloyed wells.  This is likely to be an important consideration in further experimental growth of novel quantum wells for topological insulators and other applications.

\begin{acknowledgements}
Thanks to J. Carlos Egues, P. M. Koenraad, M. B. Santos, and
S. Q. Murphy for several helpful conversations. Thanks to Craig
E. Pryor for collecting the band parameters. This project was
supported by the US National Science Foundation under
Grant~\mbox{MRSEC DMR-0080054}.
\end{acknowledgements}
\end{document}